\documentclass{article}

\usepackage{PRIMEarxiv}

\usepackage[utf8]{inputenc}
\usepackage[T1]{fontenc}
\usepackage{amsfonts}
\usepackage{amsmath}
\usepackage{array}
\usepackage{adjustbox}
\usepackage{booktabs}
\usepackage{comment}
\usepackage{csvsimple}
\usepackage{enumitem}
\usepackage{etoolbox}
\usepackage{graphicx}
\usepackage{hyperref}
\usepackage{makecell}
\usepackage{microtype}
\usepackage{multirow}
\usepackage{nicefrac}
\usepackage{pgfplots}
\usepackage{pgfplotstable}
\usepackage{tabularx}
\usepackage{url}
\usepackage{xcolor}

\hypersetup{hidelinks}

\pgfplotsset{compat=1.18}
\usetikzlibrary{patterns}

\definecolor{nord9}{HTML}{81A1C1}
\definecolor{nord14}{HTML}{A3BE8C}
\definecolor{nord11}{HTML}{BF616A}

\newcolumntype{L}[1]{>{\raggedright\arraybackslash}p{#1}}
\newcolumntype{C}[1]{>{\centering\arraybackslash}p{#1}}

\newcommand{\cmdcell}[1]{%
  \ifstrequal{#1}{POLICY_ATTESTATION (Security)}{%
    \shortstack[l]{\texttt{POLICY\_ATTESTATION}\\\texttt{(Security)}}%
  }{%
    \ifstrequal{#1}{POLICY_ATTESTATION (Consistency)}{%
      \shortstack[l]{\texttt{POLICY\_ATTESTATION}\\\texttt{(Consistency)}}%
    }{%
      \texttt{\detokenize{#1}}%
    }%
  }%
}

\newcommand{\ours}{\ensuremath{\mathsf{SxSSD}}}

\title{SxSSD: A Secure and Extensible Software-defined Solid State Drive}

\author{
  Josh Dafoe \\
  Department of Computer Science \\
  Michigan Technological University \\
  Houghton, Michigan, USA \\
  \texttt{jwdafoe@mtu.edu}
  \And
  Bo Chen \\
  Department of Computer Science \\
  Michigan Technological University \\
  Houghton, Michigan, USA \\
  \texttt{bchen@mtu.edu}
}

\begin{document}
\maketitle

\begin{abstract}

Solid-state drives (SSDs) are built on NAND flash memory and typically expose it to the operating system through a block-based storage interface. As NAND flash has special read/write constraints to accommodate its unique hardware nature, a translation between OS-level I/Os and raw flash memory I/Os is needed. This results in a flash translation layer (FTL) that essentially creates a ``trusted computing base'' due to its physical isolation from the OS. Building on this trusted computing base, some security designs (e.g., secure data recovery from malware attacks) have been proposed that can ensure strong data security properties even if the OS is compromised. However, they mostly require directly modifying the FTL's firmware code, which is hard to achieve in practice because the traditional block-based FTL does not provide an interface to modify its internal functions. New flash storage interface designs, such as open-channel SSDs or zoned namespaces, have moved key FTL functions into the OS. These new interfaces alleviate the difficulty of modifying FTL functions, at the cost of blurring the trusted boundary, as the FTL is no longer isolated from the OS. 

In this work, we have introduced SxSSD, the secure yet extensible software-defined SSD design. By decoupling the internal policy definitions from the primitive FTL mechanisms, we allow trusted applications to dynamically and securely define the FTL policies and the exposed storage interface (achieving increased flexibility compared to the open-channel and zoned namespaces SSDs). Most significantly, SxSSD can retain the isolation of traditional FTL execution (achieving a level of security similar to that of traditional block-based SSDs). We have identified and addressed a few key security challenges introduced under a compromised OS. In addition, we have implemented a prototype of SxSSD and evaluated its overhead with different FTL policies and storage interfaces. The experimental evaluation demonstrates that the overhead incurred by SxSSD is small compared to native FTL implementations.
\end{abstract}


\section{Introduction}
\label{sec:intro}
Solid-state drives (SSDs), a widely used form of data storage, are largely based on NAND flash memory. Compared to traditional magnetic storage (e.g. hard disk drives), NAND flash exhibits a few special hardware characteristics, which enforce unique low-level storage semantics. 
Specifically, data stored on NAND flash must be erased in block-sized units, typically a few hundred KiBs, while reads and writes occur in page-sized units, typically a few KiBs. In addition, updating a page requires erasing its entire containing block first, which means that data should not be overwritten in place. Therefore, a simple write request issued by the user application requires a complicated decision based on global metadata maintained throughout the flash. To give the operating system (OS) a consistent storage interface similar to traditional magnetic storage while also addressing the unique hardware constraints of flash memory, a \textit{translation} from OS-level storage requests to actual flash memory operations is necessary.

\begin{figure}[h]
    \centering
    \hspace*{0.1\linewidth}
\includegraphics[width=.45\linewidth]{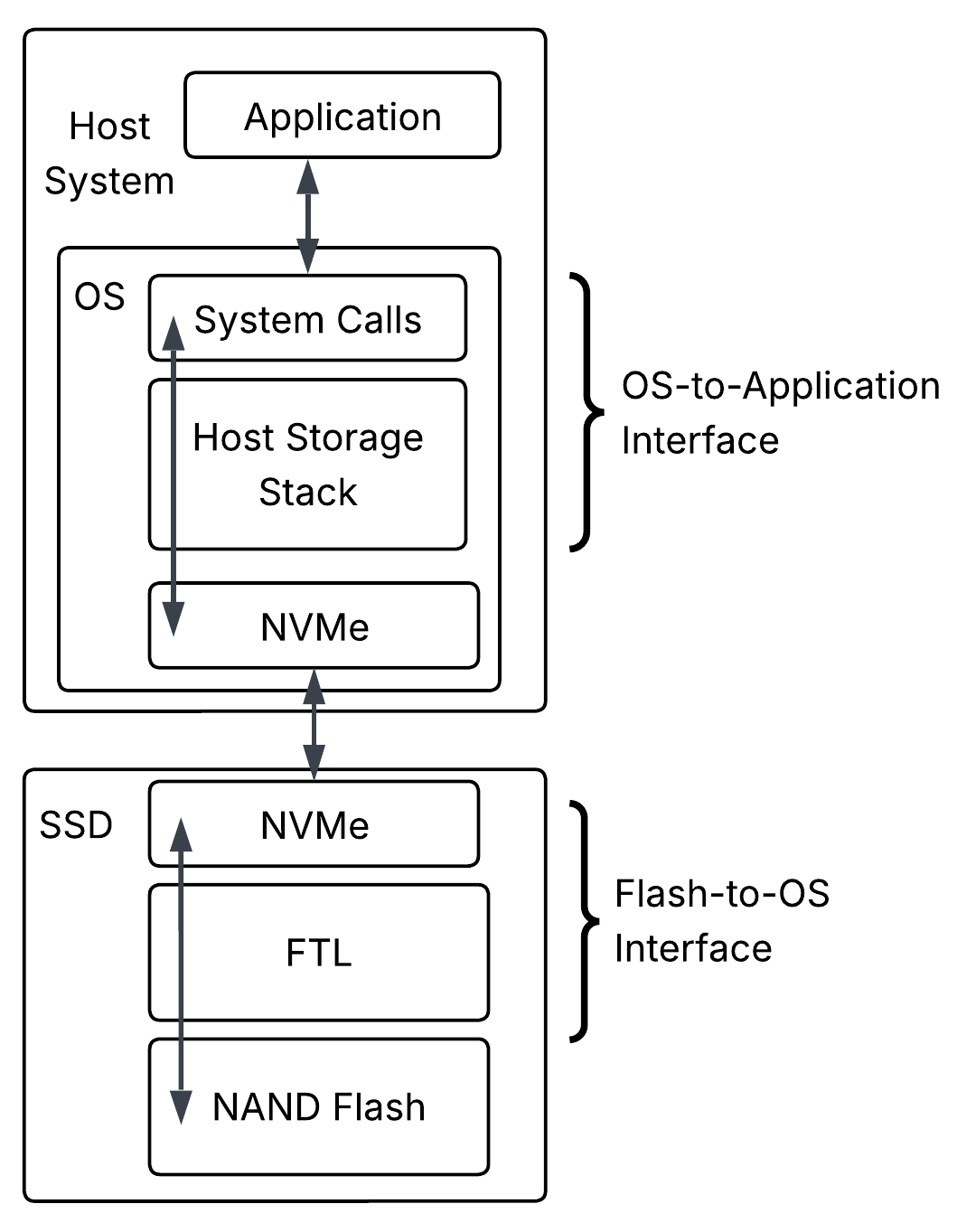}
\vspace{-5pt}
    \caption{An overview of the host and device storage stack, assuming an NVMe SSD is used. The OS-to-application interface is implemented in the host storage stack, while the flash-to-OS interface is implemented in the device-side FTL.}
    \vspace{-5pt}
    \label{fig:storage-stack}
\end{figure}

To enable this translation, a special piece of firmware, called a flash translation layer (FTL), has been introduced into modern SSDs. The FTL provides the host operating system with a logical interface (we call it \textit{flash-to-OS interface}, as shown in Figure~\ref{fig:storage-stack}) that differs from the raw flash memory interface, translating the I/O requests issued by the OS into sequences of operations supported by the underlying NAND flash. For example, a \textit{block-interface FTL} has been broadly implemented in mainstream SSDs.
The OS, in turn, often provides applications with another storage interface (we call it \textit{OS-to-application interface}, as shown in Figure~\ref{fig:storage-stack}) that is different from the one provided by the FTL. Therefore, the storage semantics ultimately observed by the applications are determined by both the flash-to-OS and OS-to-application interfaces.

Previous work has introduced software-defined SSDs (e.g., zoned namespaces (ZNS), open-channel SSD (OCSSD)) by exposing flash-to-OS interfaces that are less constrained than the traditional block interface and, therefore, closer to the underlying flash operations~\cite{nvme_zns_spec_1_3,ocssd-2018-spec,ouyang-2014-sdf,bjorling-2021-zns,stavrinos-2021-dontBeBlockhead,bjorling-2017-lightNVM}. This gives the host greater control over data placement and functions traditionally handled within the FTL, allowing the operating system to construct custom OS-to-application abstractions tailored to different workloads. As a result, applications can be co-designed with the storage stack~\cite{Picoli2020OpenChannelSSD,lerner-2021-grandpa}, and different storage semantics can be exposed to different applications~\cite{bjorling-2017-lightNVM}. 

Our key observation is that all prior SSD designs (including block-interface SSDs and software-defined SSDs such as ZNS and OCSSD) share a fundamental limitation: \textit{they expose a fixed flash-to-OS interface}. This limitation leads to two negative consequences that prevent existing FTL designs from meeting the diverse requirements of emerging applications:

\noindent \textbf{(1) Difficulty in Building High-performance Storage Systems.} 
Previous work has consistently demonstrated that flash storage performance depends strongly on the dynamic interaction between specific workloads and the underlying flash management strategies~\cite{agrawal-2008-designTradeoffs,he-2017-unwritten,bjorling-2021-zns}. However, implementing this ``smart'' interaction is hard in existing FTL designs, as they all expose a fixed flash-to-OS interface. For block-interface SSDs, the OS typically interacts with the FTL via a limited block-access interface, making it difficult to coordinate the application workloads with the FTL. For example, many applications may suffer from the log on log problem~\cite{yang-2014-logOnLog}, resulting in unnecessary duplicate functions between FTL and the upper-layer software~\cite{yang-2014-logOnLog}. 
Software-defined SSDs such as ZNS and OCSSD may partially address this problem by introducing programmability of the OS-to-application interface. However, since their flash-to-OS interface is still fixed, they remain constrained by a single set of storage semantics that inherently favors some workloads over others. For example, ZNS can make some host-side operations (e.g. zone compaction) more expensive than necessary by forcing them through a limited flash-to-OS interface~\cite{tan-2024-optimizing, han-2021-zns+}. Overall, although many applications obtain performance benefits from modifying the device-level FTL~\cite{han-2021-zns+,jiao-2022-wearHarm,Chen-2011-CAFTL,kim-2015-SLOComplyingSSD,huang-2017-flashblox,ssd-batch-2021}, the fixed flash-to-OS interface makes such modifications difficult in practice.

\noindent\textbf{(2) Difficulty in supporting critical security storage semantics.} The device-side FTL is isolated from the host OS by SSD hardware and is therefore a critical security boundary when facing a compromised OS~\cite{huang-2017-flashguard,baek-2018-ssdInsider}. Previous work has leveraged this by designing modified block-interface FTLs to provide security-critical storage semantics to trusted applications, such as plausibly deniable storage~\cite{chen-2021-plausibleDeniability}, secure data recovery from malware attacks~\cite{zhu-2025-last,baek-2018-ssdInsider,huang-2017-flashguard,guan2017supporting}, integrity-assured cloud storage~\cite{dafoe-2025-hardware}, etc. Since these modifications rely on the trustworthy nature of the device-side FTL, implementing these designs currently requires explicit modifications to the FTL firmware. Thus, it is difficult (or even impossible) for existing SSDs (including both block-interface and software-defined SSDs) to support such security-critical designs due to the inflexible nature of the flash-to-OS interface in existing FTLs. Notably, the existing software-defined SSDs do not address this problem: While they introduce custom storage semantics at the OS-to-application interface, they fail to support the class of security-critical storage semantics because the flash-to-OS interface remains fixed.

Until now, there has been no existing approach that can overcome the fundamental limitation mentioned above. In this work, we introduce $\ours$, the \textbf{S}ecure yet e\textbf{x}tensible software-defined \textbf{S}olid \textbf{S}tate \textbf{D}rive, aiming to close this gap. Instead of introducing another FTL with a fixed flash-to-OS interface, $\ours$ proposes to introduce extensibility on top of a set of generic FTL procedures by allowing the flash-to-OS storage interface and the supporting FTL policies to be dynamically defined while maintaining the trusted device boundary. Our design is guided by two central requirements: extensibility and secure policy management:

\noindent\textbf{Extensibility}. We address the fundamental limitation of a fixed flash-to-OS interface through extensibility, which allows trusted applications to define new storage semantics and FTL behavior. Specifically, $\ours$ achieves this extensibility through programmability, by allowing custom policy code and storage semantics to be defined and installed by trusted applications without modifying the FTL firmware. This addresses the difficulty in building high-performance storage systems by allowing FTL behavior to be dynamically tailored to specific workloads, rather than being constrained by a fixed interface. Additionally, this extensibility partially addresses the difficulty of supporting security-critical storage semantics, since such semantics are impractical to realize within the trusted device boundary without a dynamically definable flash-to-OS interface.
To enable extensibility, the \textit{first challenge} is to establish a general programming model to define custom flash-to-OS storage semantics. To address this, we characterize the FTL design space and identify a common structure in which fundamental FTL mechanisms are clearly distinguished from the policies that determine flash-to-OS storage semantics (Section~\ref{sec:characterization}). The \textit{second challenge} is to provide a concrete design through which this programmability can be realized. To address this, the design of $\ours$ is informed by the established mechanism-policy decomposition. Specifically, $\ours$ is structured so that the core flash management mechanisms are fixed. Then, on top of these, $\ours$ exposes a well-defined API (to FTL policies) that allows interaction with the underlying storage hardware, through which they realize host-visible storage semantics. To ensure that custom policies are enforced on-device, policies are (1) stored in flash storage, (2) dynamically loaded into memory, and (3) invoked through an event-driven policy engine. Finally, to expose the programmability externally, $\ours$ introduces an interface for installing and managing custom device-enforced policy code. We call this new interface a \textit{meta-interface}, since it operates at a higher level of abstraction, enabling the definition of custom storage interfaces (i.e. it is ``an interface for defining interfaces''), rather than providing fixed storage semantics. Unfortunately, expanding the FTL interface in this way introduces a new attack surface through which an untrusted host OS may attempt to manipulate policies. Our next requirement effectively removes this attack surface by restricting access to the meta-interface.

\noindent\textbf{Secure Policy Management}. This requirement enables security-critical semantics by ensuring that policies can only be managed by trusted entities, even with an untrusted OS as a mediator. The \textit{first challenge} is to prevent untrusted entities from modifying functionality beneath the trusted device boundary. To address this, $\ours$ makes the meta-interface the exclusive mechanism through which policies can be managed. In addition, we restrict its use to trusted entities through authentication mechanisms. The \textit{second challenge} is to ensure that the policy integrity is maintained, and the on-device policy state can be audited. To address this, $\ours$ supports policy attestation and protects policy integrity by storing policies in a reserved on-device region outside the policy-visible address space. When required by the trusted application, $\ours$ also protects the confidentiality of the policy during management and storage, preventing the adversarial host from inferring the semantics of the policy.

\noindent\textbf{Contributions:} Our contributions are as follows:
\begin{enumerate}[label=(\roman*)]
    \item We survey existing storage interface designs and characterize the FTL design space under a set of unifying abstractions.
    \item We propose $\ours$, which restructures the FTL according to this characterization and exposes a meta-interface for dynamically defining host-visible storage semantics and installing corresponding on-device policies.
    \item We introduce mechanisms to make extensibility secure under an untrusted host OS, including authorized policy management, policy attestation, policy freshness, policy integrity, and policy confidentiality.
    \item We have implemented a prototype of $\ours$ in an SSD emulator~\cite{li-2018-femu} and evaluated it on multiple real-world workloads, showing only 3.74\% overhead over a baseline SSD.
    \item We conduct a case study in which we enable secure data recovery from ransomware attacks, demonstrating $\ours$ can support various applications without the need to modify the FTL firmware code. 
\end{enumerate}

\section{Background}
\label{sec:background}
\subsection{NAND Flash Characteristics}
\label{sec:background:nand}
NAND flash presents hardware characteristics that differ from magnetic storage. It uses floating gate transistors, which are organized onto a number of storage chips. These are then wired to a processor via independent channels, allowing parallelism across these channels. The storage chips are organized into a hierarchy of die, plane, block, and page. While there are multiple dies per channel, each die allows only a single flash command to be executed at a time. However, that command can be executed synchronously at aligned addresses on multiple planes in parallel. Within each plane, NAND is organized into blocks and pages; each plane contains the same number of blocks, and each block contains the same number of pages. Importantly, the parallelism exhibited by flash memory can significantly increase its performance. Therefore, the optimal strategy is to stripe the writes across the various units of parallelism (Figure~\ref{fig:nand-geometry}), rather than writing sequentially within a single unit. Finally, each physical page includes a small out-of-band (OOB) area that is used internally by the FTL.

Due to the nature of floating gate transistors, there are a few constraints on the primitive read/write/erase operations: (\textbf{C1}) Writes are performed at page granularity. (\textbf{C2}) Erase operations (which are an order of magnitude more expensive than write) must be performed before overwriting a page. Additionally, primitive erase operations are performed only at the block granularity. (\textbf{C3}) Writes are often constrained to proceed sequentially within a NAND flash block. (\textbf{C4}) Flash blocks can support only a limited number of erasures before failure and therefore have a limited lifetime~\cite{bonnet-2011-dbms}.

Together, these constraints characterize the raw storage interface exposed by NAND flash hardware. However, C1-C4 result in complex dependencies between flash operations over time, making it challenging to present a simple unified interface that effectively virtualizes storage resources among applications. Instead, I/O requests issued by higher software layers must be translated into sequences of flash operations that satisfy C1-C4 while preserving the expected storage semantics. Although such translation may occur at multiple layers in the storage stack, in this paper, we use the term \textit{flash translation layer} (FTL) to refer specifically to the translation software running on the storage controller.

Typically, flash based SSDs are equipped with DRAM sufficient for a read/write cache and efficient FTL execution. Often, this DRAM can resist power loss failure by incorporating capacitors~\cite{pott2014supercapacitors} or batteries~\cite{kateja-2017-battery}.

\subsection{The NVMe Interface}
\label{sec:background:nvme}

Mainstream SSD interface types include Serial ATA (SATA), Serial Attached SCSI (SAS), and NVM Express (NVMe), among which NVMe SSDs are expected to have the fastest average yearly growth rate~\cite{foretune-report-ssd-2026}. For NVMe SSDs, communication between the host OS and the SSDs typically takes place using the NVMe over PCIe protocol (Figure~\ref{fig:storage-stack}). This interface relies on submission queues and completion queues resident on the host DRAM. The host invokes storage operations by placing NVMe commands into a submission queue, then the FTL fetches these commands, processes them, and reports this via a completion queue. Two aspects of the NVMe interface are particularly relevant to our design: First, NVMe simply presents the device address space as a flat logical block addressing scheme, effectively decoupling the host-side view from the internal geometry managed by the FTL. This implies that some translation into the physical address space is inherent to all FTLs. Second, the NVMe command set supports vendor-specific commands in addition to a set of standard operations (e.g., read, write).  We leverage and expand this flexibility to define our own NVMe commands and enable policy-specific NVMe commands.

\subsection{Case Studies of Flash Translation Layers}
\label{sec:background:case-studies}

\begin{figure}[t]
    \centering
    \includegraphics[width=.45\linewidth]{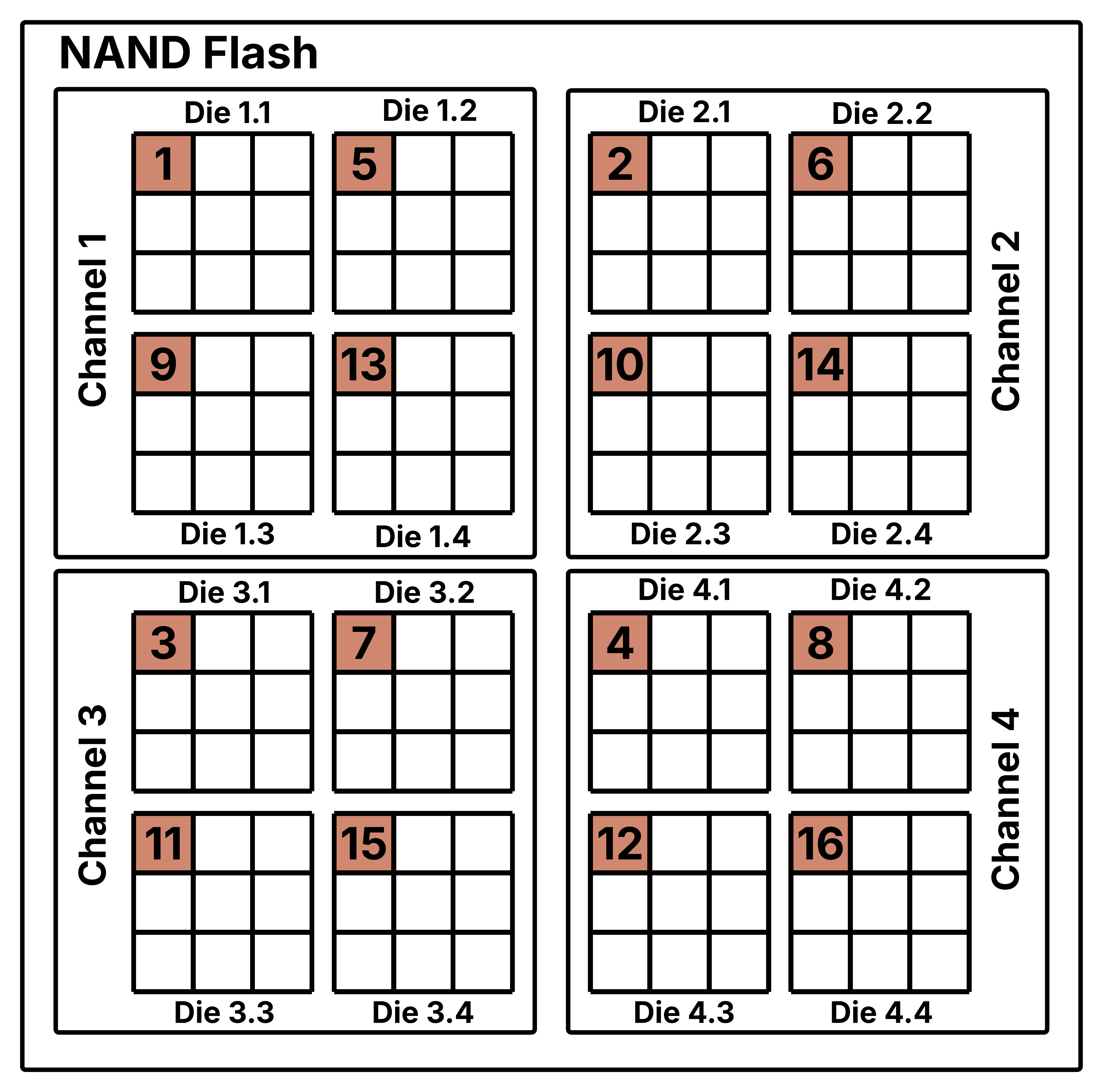}
    \caption{A simplified view of NAND flash geometry. The orange blocks make up a superblock, commonly used by block-interface and ZNS FTLs. The numbers indicate the striping order that maximizes parallelism across the superblock.}
    \vspace{-10pt}
    \label{fig:nand-geometry}
\end{figure}

In the following, we survey the main FTL designs, in order to establish a unifying characterization of the FTL design space, which informs our subsequent conceptual separation between FTL mechanisms and policy (Section~\ref{sec:characterization}).

\subsubsection{Open Channel SSDs}

An open-channel SSD (\textbf{OCSSD}) exposes internal flash structure and parallelism to the host through its FTL, presenting the OS with an interface closely matched to raw flash semantics. 
Specifically, the OCSSD presents a logical address space with a fixed mapping onto the device geometry that reflects hardware parallelism. As a consequence, the OS can affect performance via direct control over data placement.

Complying with C1, an OCSSD defines the smallest addressable unit for read/write as a logical block, often corresponding to a physical page. Logical blocks are organized into chunks, each associated with a parallel unit (PU), with PUs organized into groups. Often, groups correspond to channels, and PUs correspond to dies. Writes must advance sequentially within chunks along an FTL-maintained write pointer. To exploit parallelism within a chunk, the write pointer advances at the same page index across different planes. Then, C3 is addressed by advancing the write pointer to the next page index after exhausting the planes. This implies that each chunk consists of a fixed set of aligned physical blocks within its associated PU, so each physical block belongs to exactly one chunk.
In addition to a write pointer, the OCSSD FTL maintains chunk state information (free, open, closed, or offline). To address C2, the enforced erase unit is a chunk (i.e., a fixed set of aligned physical blocks within a PU), which must be reset before it can be rewritten. In response to C4, the OCSSD FTL tracks the wear status of the chunk and reports this to the host, which is responsible for implementing wear leveling policies to evenly distribute wear between chunks. Malfunctioning chunks are retired in the FTL by internally setting them to the offline state~\cite{ocssd-2018-spec}.

\subsubsection{Zoned Namespaces SSDs}
Rather than exposing the raw flash geometry and device parallelism to the host, the zoned namespaces (\textbf{ZNS}) interface simply exposes logical zones with a sequential write constraint. Although zones are fixed logical regions of the namespace, their placement on physical media is managed internally by the ZNS FTL and may change over time. Although implementation details can vary, zones typically span aligned block addresses across multiple dies, with the write pointer advancing along aligned page offsets in those blocks. A zone may even cover an entire superblock, which is a common block address that spans all channels, dies, and planes (Figure~\ref{fig:nand-geometry}).
Constraints C1 and C3 are addressed similarly to the OCSSD. Specifically, logical blocks are the smallest unit of read/write, and sequential writes within zones ensure that the write pointer advances without violating C3. Additionally, the ZNS FTL maintains a per-zone write pointer and tracks zone state (empty, open, closed, full, read-only, or offline). To address C2, the enforced unit of erase is a zone, which must be reset before it can be rewritten. Unlike OCSSD, C4 is addressed by performing wear-leveling in the FTL, in which physical resources are dynamically allocated to zones to spread erase counts evenly across blocks~\cite{Liu2024ZNSCleanerEL}. In addition, zones are retired after failures by moving them into the offline state. Importantly, the ZNS SSD is considered a successor to OCSSD~\cite{bjorling-2021-zns, stavrinos-2021-dontBeBlockhead}, and can be implemented on top of the OCSSD interface~\cite{Picoli2020OpenChannelSSD}.

\subsubsection{Block-interface SSDs}
\label{sec:block-SSD}

Block-interface FTLs hide the flash storage constraints from the host. This is achieved by presenting a standard logical block interface, enabling random read/write operations over the entire logical address space. For C1, the unit of read/write is maintained as a logical page, corresponding to the size of a physical page. To handle C2, the FTL implements out-of-place updates, where logical overwrites are written at different physical locations. Typically, the new writes are striped along superblocks (Figure~\ref{fig:nand-geometry}), to maximize internal parallelism (thereby satisfying C3 similarly to OCSSD and ZNS).
When this occurs, the old physical locations are invalidated. Invalid pages accumulate over time and must be reclaimed by on-device garbage collection (typically at the superblock level). In addition, to keep track of the evolving physical location of a given logical page, a mapping table is maintained.

To address C4, the block-interface FTL performs internal wear leveling to balance erase counts across blocks.
Usually, this is done by migrating blocks (or superblocks) that have divergent erasure counts~\cite{chang-2007-dpwl}. Additionally, bad block management is performed on-device. Retired blocks are hidden via an indirect layer between the raw flash blocks and pseudo physical blocks~\cite{choi-2018-hil,kim-2017-pbms}. For a given pseudo physical block address, the mapping is an identity mapping if the corresponding physical block is intact, but when a physical block is retired, the corresponding pseudo physical block is remapped to a spare block from an overprovisioned pool.

Importantly, previous work has shown that a block abstraction can be implemented on top of the ZNS interface~\cite{bjorling-2021-zns, holmberg_dmzap_2021, zonedstorage_kernel_config} or OCSSD interface~\cite{bjorling-2017-lightNVM}. Also, ZNS can be built on top of OCSSD~\cite{Picoli2020OpenChannelSSD}. This suggests that an FTL implementing the functionality necessary to expose a block interface can be viewed as implicitly containing the internal machinery that ZNS and OCSSD expose explicitly. From this observation, we arrive at a general question, which we address in Section~\ref{sec:characterization}: can we characterize the FTL design space under a framework that cleanly separates FTL mechanisms and flash management policies, so that different storage semantics (e.g. OCSSD, ZNS, block interface) and the policies can be exposed dynamically?

\section{Models and Assumptions}
\label{sec:model}

We consider a computing device equipped with a NAND-based NVMe SSD (Figure~\ref{fig:storage-stack}). We assume that the SSD correctly runs the firmware that implements our $\ours$ design. This assumption is reasonable because 1) the independent flash storage hardware (processor, memory) provides isolation between the FTL firmware and the host OS, and 2) methods such as secure boot~\cite{wd_drive_protection, micron_ssd_protecting_your_ssd_and_your_data_2016} can prevent unauthorized firmware modification. 
We do not consider data loss in the DRAM cache due to power failure as modern SSDs can be equipped with capacitor or battery backed DRAM~\cite{pott2014supercapacitors,kateja-2017-battery}.

We consider that the computing device runs trusted applications controlled by a trusted system administrator. The integrity and isolation of these applications can be maintained by running them in a secure environment such as a trusted execution environment~\cite{dafoe-2025-hardware,liao-2024-fspde,ahn-2020-diskshield}, secure bootloader~\cite{Chen2022ACP}, or remote trusted computer~\cite{wang-2019-almanac}. These mechanisms can prevent applications from being tampered with even in the presence of a compromised operating system.
We assume that all entities authorized to manage policies are mutually trusted. 

We assume that the host OS can be compromised by an adversary capable of performing privilege escalation and obtaining root access. Hence, the host OS is untrusted. We model this as a Dolev-Yao-style adversary which controls the communication channel between the trusted application and the interface exposed by the SSD, yet cannot feasibly break cryptographic primitives~\cite{dolev-1983-security}. This implies that the host OS has full access to the interface exposed by the SSD, at any time. In particular, the adversary may be active while the system administrator is accessing the policy management interface (i.e. meta-interface).

We also assume that the SSD contains a device-specific key-pair, 
generated and embedded by the manufacturer, as assumed in previous work~\cite{ahn-2020-diskshield, zhu-2025-last}. A certificate 
associated with this key is provided by the manufacturer to the trusted system administrator. 
In addition, the system administrator generates a key-pair
before obtaining the SSD. Through a secure communication channel with the manufacturer, the administrator's public key is certified. The resulting certificate 
is provisioned to the SSD.
Importantly, this grants the system administrator the ability to install custom policies onto $\ours$, as detailed in \S~\ref{sec:design:meta}. The system administrator directly controls a trusted application that can (1) install policies onto $\ours$, and (2) interact with those policies through their exposed interfaces.

\section{A Characterization of the FTL Design Space}
\label{sec:characterization}

In this section, we answer the question posed at the end of Section~\ref{sec:background:case-studies}. We identify common structural properties across the major FTL designs and examine how different FTL designs expose these properties to the OS (Section~\ref{sec:characterization:structure}). Based on the observations made in Section~\ref{sec:characterization:structure}, we define a new conceptual model for our $\ours$ design (Section~\ref{sec:characterization:model}). This new model allows trusted applications to dynamically define FTL policies so that the flash-to-OS interface turns changeable. 

\subsection{Common Structural Properties of FTLs}
\label{sec:characterization:structure}
\subsubsection{Internal Parallelism and Sequential Write Domains}

Our first observation is that FTLs internally stripe writes across some hierarchy level (i.e. channel, die, plane, block, or page) to exploit device-level parallelism. Additionally, we observe that block-interface, open-channel, and ZNS SSDs manage the internal striping through sequential-write constraints, but differ in (1) the size of the ``striping domain'', (2) the striping level, and (3) how they expose this to the host. Illustrating (1) and (2), as discussed in Section~\ref{sec:background:case-studies}, OCSSDs typically write data in chunks, striping successive writes across planes within a die, whereas ZNS and block-interface SSDs typically write into superblocks that stripe successive writes across channels first (see Figure~\ref{fig:nand-geometry}). Then, demonstrating (3), block-interface SSDs hide the sequential-write constraint behind an additional layer of indirection through a page-level mapping table.

To make this precise, we define a logical address space as any address space constructed inside the FTL above raw physical flash, independent of whether it is exposed to the host. A striping domain is a group of addressable units in an address space across which the FTL advances writes to exploit internal parallelism, with writes ultimately striped over an underlying group of physical pages (e.g. a superblock). A sequential write domain (\textbf{SWD}) is the logical view of a striping domain: writes are sequential with respect to the SWD's logical write pointer, even though the underlying striping domain may place consecutive logical writes across multiple channels, dies, and/or planes. For example, a ZNS zone can be viewed as an SWD with a superblock as its striping domain. 

Given a striping level and SWD size, an address space can be partitioned into SWDs. At the lowest level, the raw flash consists of physical blocks. On top of this, bad block management (BBM) introduces a pseudo-physical address space by mapping pseudo-physical blocks to physical blocks. Each pseudo-physical block is then a striping domain with page-level striping. Since an SWD is the logical view of a striping domain, each pseudo-physical block has a corresponding ``primitive'' SWD (we call these \textit{pSWDs}).
Higher-level SWDs exposed to FTL policy (we call these \textit{eSWDs}) may then be composed of multiple pSWDs. Here, \textit{exposed to policy} refers to being visible to the internal FTL policy logic, independent of how the policy exposes this through the host interface.

From the above, we see that OCSSDs expose SWDs composed of $planes\_per\_die$ pSWDs with plane-level striping, whereas ZNS and block-interface FTLs use SWDs that span $planes\_per\_die$ $\times$ $dies\_per\_channel$ $\times$ $num\_channels$ pSWDs with channel-level striping (i.e. superblocks). The key difference between ZNS and block-interface FTLs is how these eSWDs are exposed \textit{to the host}: ZNS exposes SWDs directly to the host as zones, while block-interface FTLs instead present a flat logical address space and map host writes onto these eSWDs through a page-level mapping table.

Across all designs, SWDs are associated with state machines that maintain internal state, such as mapping, wear status, write pointers, and SWD state (e.g., free, open, closed). For SWDs exposed to policy, this state can be extended with policy-defined state to realize particular semantics. A subset of the resulting state, selected by policy and relevant to the interface, may become host-visible, while the remainder stays device-internal.

\subsubsection{Event-Driven Policies}
\label{sec:characterization:event}

Our second observation is that all policy-level actions are naturally event driven, because the normal FTL functions (e.g. mapping, garbage collection, wear leveling, and bad block management (BBM)) take place in response to changes in the device state. That is, policies are triggered when certain conditions over state maintained by the per-SWD state machines are satisfied. Additionally, this state changes during specific ``events'' such as writing a new page, erasing a block, etc. As an example, the per physical block wear status is tracked by the associated pSWD state machine. Wear leveling (e.g., included in ZNS and block interface SSDs) is then triggered when erase counts diverge beyond some threshold, with this condition evaluated on write, erase, or background events. On the other hand, OCSSDs simply report 
wear status to the OS which performs wear leveling manually. 
Additionally, as mentioned above (\S~\ref{sec:block-SSD}), BBM policies respond to physical block failure events by either remapping the affected pseudo-physical block to a healthy block, or retiring it. We can model retirement (used by the OCSSD and ZNS BBM policy) as a degenerate case of remapping, in which the pseudo physical block is remapped to null. Together, these examples illustrate that \textit{diverse FTL behaviors can be modeled as event-driven actions triggered by conditions over the SWD state}. This naturally extends to other FTL functions such as garbage collection and address translation.

\subsection{A Conceptual Model of our New FTL Design}
\label{sec:characterization:model}

The structural properties identified in Section~\ref{sec:characterization:structure} suggest that FTL designs share a common underlying organization where writes advance over SWDs with associated state machines, and policy-level behavior is triggered by events over this state. Based on this key observation, we define a new conceptual model for the FTL design.
Specifically, we can model an FTL as being composed of five conceptual components: \textbf{(M1)} the raw flash storage, consisting of the physical geometry constrained by C1-C4; \textbf{(M2a)} a mechanism layer that constructs a pseudo-physical address space through bad block management (BBM) and defines the primitive SWDs (\textit{pSWDs}) over that space; \textbf{(M2b)} a mechanism layer that groups pSWDs into higher-level SWDs exposed to policy (\textit{eSWDs}); \textbf{(M3)} a set of events, triggered either internally or by specific NVMe commands; \textbf{(M4)} the policy-defined eSWD geometry, context, and state machines, and \textbf{(M5)} a set of policies defined as pairs of (condition, action) functions attached to events.

To make the FTL extensible, we fix M1-M3 as mechanisms, and add the meta-interface, which makes M4-M5 (policy) programmable, enabling customization of the host-visible interface semantics. We expand on the definitions of M1-M5 below:

Similarly to \textbf{(M1)}, any FTL is built on top of raw flash storage hardware that exhibits constraints C1-C4. Thus, M1 consists of the read/write/erase operations consistent with these constraints.

Corresponding to \textbf{(M2a)}, the BBM indirection layer introduces a \textit{pseudo-physical} address space, where each pseudo-physical block is a block-sized striping domain with page-level striping. Then, each pseudo-physical block creates a pSWD (thus, the striping domain of a pSWD is a pseudo-physical block). BBM then maintains a mapping from pseudo-physical to physical blocks, with overprovisioned physical blocks initially left unmapped. The BBM layer exposes an API to policies for maintaining the pseudo-physical to physical block mapping over time, and migrating valid data when a pseudo-physical block must be remapped to a new physical block. As established in Section~\ref{sec:characterization:event}, this model is sufficient to implement the policies of ZNS, OCSSD, and block-interface FTLs.  

\noindent\fbox{%
  \begin{minipage}{\dimexpr\linewidth-2\fboxsep-2\fboxrule\relax}
  \textbf{pSWD state and context.}
  \begin{itemize}
      \renewcommand\labelitemi{$-$}
      \item \textbf{Mapping:} The physical block currently mapped to the underlying pseudo-physical block.
      \item \textbf{Wear status:} Erase count of the mapped physical block.
      \item \textbf{pSWD state:} Free, open, closed.
      \item \textbf{Write pointer:} The next page to be written.
      \item \textbf{Page validity:} Per-page status (free, valid, or invalid).
  \end{itemize}
  \end{minipage}%
  }

Corresponding to \textbf{(M2b)}, the next mechanism layer is a higher-level abstraction over the pSWD space to policies. This abstraction, called an exposed SWD (eSWD), groups pSWDs into policy-visible units over which writes are striped according to the striping domain and striping level selected by the policy.

The API exposed to policy for managing eSWDs provides four key capabilities: 1) accessing and changing the mapping between pSWDs and eSWDs, 2) migrating data between eSWDs, 3) issuing read, sequential write, and erase operations over eSWDs, and 4) managing the eSWD context.

  \noindent\fbox{%
  \begin{minipage}{\dimexpr\linewidth-2\fboxsep-2\fboxrule\relax}
  \textbf{Mechanism-level eSWD context.}
  \begin{itemize}
      \renewcommand\labelitemi{$-$}
      \item \textbf{Mapping:} A mapping to the pSWDs covered.
      \item \textbf{Layout:} The policy-defined eSWD size and striping level.
      \item \textbf{Write pointer:} The next page to be written.
  \end{itemize}
  \end{minipage}%
  }

The FTL events \textbf{(M3)} trigger specific policy functions when they occur and so are the mechanisms through which policies are called.

  \noindent\fbox{%
  \begin{minipage}{\dimexpr\linewidth-2\fboxsep-2\fboxrule\relax}
  \textbf{FTL event types (M3).}
  \begin{itemize}
      \renewcommand\labelitemi{$-$}
      \item The background processing event.
      \item NVMe commands.
      \item pSWD state transitions.
      \item Error events over primitive flash operations.
  \end{itemize}
  \end{minipage}%
  }

For \textbf{(M4)}, policy defines the logical organization and semantics of eSWDs. As mentioned above, this includes defining the size and striping level of eSWDs, which determines how M2b maps logical writes onto the underlying pseudo-physical space. Policy also defines any additional eSWD-level context and state machines needed to realize a particular flash-to-OS interface. In this way, M4 captures the storage semantics that distinguish one FTL interface from another. 

For example, a block-interface policy uses eSWDs spanning superblocks with channel-level striping. Additionally, this policy maintains a page-level mapping table. On NVMe write events, the block-interface policy would write the new page at the write pointer of the current open eSWD, update the mapping table to point to the new physical location, and invalidate the old physical page.

Then, corresponding to \textbf{(M5)}, policies are expressed as (condition, action) pairs stored in function pointer tables attached to events. Conditions are boolean functions defined in terms of policy specific data structures, state-machine status, and event-specific information. When a condition is satisfied, the corresponding action function is executed. The action interacts with flash storage through the policy-visible API (M2).

\section{\texorpdfstring{$\ours$}{SxSSD} Design}
\label{sec:design}

\subsection{Design Overview}

\begin{figure}[t]
      \centering
      \includegraphics[width=.5\linewidth]{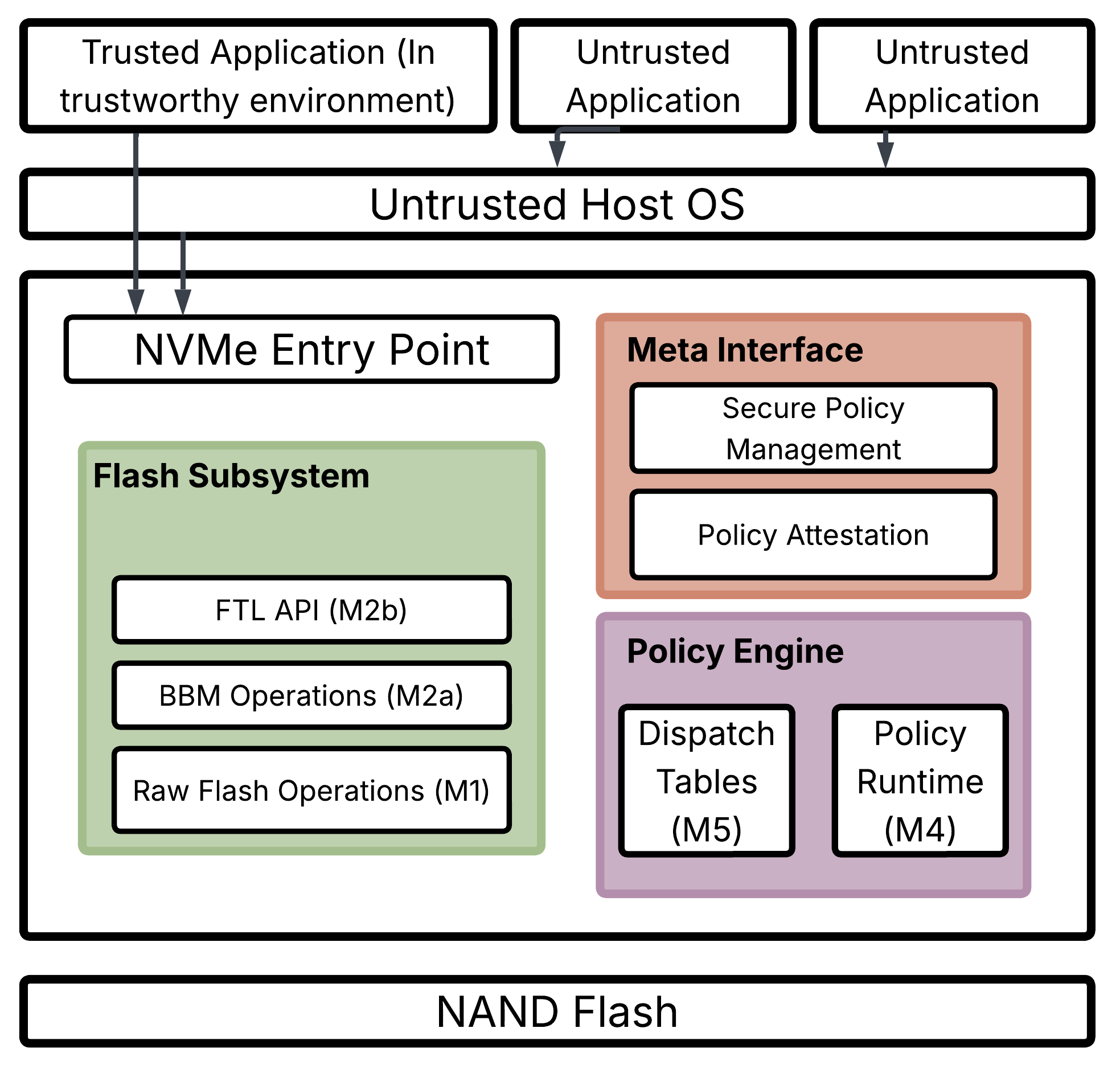}
      \vspace{-5pt}
      \caption{An overview of $\ours$. Note that events (M3) are generated across multiple components.}
      \vspace{-10pt}
      \label{fig:design}
  \end{figure}

To enable programmability of both the host-visible storage interface exposed by $\ours$ and the on-device policies that implement it, we propose a new FTL structure that realizes the conceptual FTL model defined in Section~\ref{sec:characterization:model}. Our resulting FTL structure consists of three main components, shown in Figure~\ref{fig:design}: 
(1) the \textbf{flash subsystem}, which instantiates M1-M2 and presents an API for policy code to call into; (2) the \textbf{policy engine}, which manages runtime policy state and context (M4), and executes policy code by routing events (M3) to the policy functions (M5) via dispatch tables; and (3) the \textbf{meta-interface}, which enables secure management of the policies that implement M4 and M5. 

The meta-interface provides mechanisms for the upper layer to manage custom policies on the FTL. This includes installing, updating, temporarily deactivating, and removing policies. However, exposing this interface to the host OS without carefully restricting how upper-layer software may invoke it creates a critical attack surface. Accordingly, a significant responsibility of the meta-interface includes restricting access to policy management and ensuring that only the trusted application(s) can manage policies, even in the presence of an actively adversarial host OS with Dolev-Yao capabilities. Details of the meta-interface are elaborated in Section~\ref{sec:design:meta}.
 
After a policy is stored, it must be loaded and enforced at runtime. The policy engine contains the policy storage and the mechanisms for loading policies, maintaining their state, and calling into them when events are triggered. Its core component is a set of dispatch tables, each associated with an event, that point to the corresponding policy handlers (i.e. (condition, action) pairs). Corresponding to M3, the events include receiving NVMe commands, state transitions in the primitive SWD state machines, errors on primitive flash operations, and the background event. The details of the policy engine are elaborated in Section~\ref{sec:policy-engine}.

To enable interaction between policy functions and flash storage, the flash subsystem implements the storage management mechanisms and exposes them to policy through a well-defined interface. We construct this subsystem by starting from primitive flash operations constrained by (C1-C4), and adding a few powerful abstractions on top: The raw flash operations are the lowest layer, performing read/write/erase on individual pages and blocks. The BBM indirection layer is implemented on top of the raw flash operations, while the eSWD abstraction and policy API are built on top of the BBM primitives. It is expected that regular policies only call into the policy API (Section~\ref{sec:discussion}). The details of the flash subsystem are elaborated in Section~\ref{sec:flash-subsystem}.

\subsection{The Meta-Interface}
\label{sec:design:meta}

The meta-interface provides the host with a small set of primitives for securely managing on-device policies. These functionalities are presented to the host as new vendor-specific NVMe commands, described in Table~\ref{tab:meta_nvme_cmds}. Since each NVMe command triggers an event that indexes into the policy engine's dispatch table, the meta-interface is itself implemented as a privileged policy.
To prevent an untrusted host from abusing this interface, the meta-interface policy should authenticate requests and ensure their integrity and freshness.  

  \begin{table}[tb]
  \centering
  \setlength{\tabcolsep}{6pt}
  \begin{tabular}{p{0.35\linewidth} p{0.55\linewidth}}
  \toprule
  \textbf{NVMe Command} & \textbf{Description} \\
  \midrule
  \texttt{INIT\_SESSION} & \emph{Resets the counter ($\mathsf{ctr}_0$), establishes a session key ($K_s$) shared between a trusted application and $\ours$, and sets the session mode
  (normal or confidential).} \\
  \midrule
  \texttt{INSTALL\_POLICY} & \emph{Installs a new policy into the overprovisioned policy storage area, and tracks relevant metadata.} \\
  \midrule
  \texttt{UPDATE\_POLICY} & \emph{Replaces the code of an existing policy while preserving its policy ID.} \\
  \midrule
  \texttt{REMOVE\_POLICY} & \emph{Deletes a policy and its associated metadata from the overprovisioned storage area.} \\
  \midrule
  \texttt{ACTIVATE\_POLICY} & \emph{Activates a policy by reading it, parsing its code, then loading it into the policy engine.} \\
  \midrule
  \texttt{DEACTIVATE\_POLICY} & \emph{Deactivates a policy by removing its functions from the policy engine.} \\
  \midrule
  \texttt{POLICY\_ATTESTATION} & \emph{Generates an authenticated attestation report on a policy's installation status, activation state, and optionally its code integrity via hash
  verification.} \\
  \bottomrule
  \end{tabular}
  \caption{Vendor-specific NVMe commands introduced by the $\ours$ meta-interface. All meta-interface commands (except \texttt{INIT\_SESSION}) require authentication with $K_s$ and
  include counter-based replay protection. In confidential mode, request data are encrypted.}
  \vspace{-20pt}
  \label{tab:meta_nvme_cmds}
  \end{table}

\subsubsection{Authorized Policy Management with Replay Protection}

Towards achieving this, we add an NVMe command \texttt{INIT\_SESSION} that establishes a session between a trusted application and the SSD. Part of this process is to generate a shared session key $K_s$. Because both parties possess each other's certificates, this can be done securely using
an authenticated key exchange protocol~\cite{Diffie1992AuthenticationAA}.

To secure the policy management NVMe commands, the meta-interface only accepts requests generated by entities holding $K_s$ (i.e. a trusted application).
This is enforced by authenticating each request using a message authentication code under key $K_s$. Importantly, this enables the FTL to check both the source and integrity of any request data that are covered by the MAC. The attacker might manipulate the NVMe opcode itself, so that both the opcode and request data should be authenticated. As an initial solution, before calling the NVMe commands (Table~\ref{tab:meta_nvme_cmds}, excluding \texttt{INIT\_SESSION}), the trusted application must compute a MAC over the NVMe opcode concatenated with the request data. The FTL can then verify the MAC correctness before continuing to handle the relevant NVMe command. 

Unfortunately, this method does not prevent an adversary from replaying valid policy management requests. To prevent this, we introduce a counter $\mathsf{ctr}_0$ maintained by the meta-interface that is reset whenever a call to \texttt{INIT\_SESSION} succeeds. Correspondingly, the trusted application maintains a counter $\mathsf{ctr}_1$. Then, on any call into the meta-interface, the trusted application will include $\mathsf{ctr}_1$ in the request data that is covered by the MAC. In turn, the FTL enforces that $\mathsf{ctr}_1 > \mathsf{ctr}_0$ before authenticating the request, and sets $\mathsf{ctr}_0 \leftarrow \mathsf{ctr}_1$ afterwards. This ensures that even if the original contents of a request are identical, either the session key or the counter value included in the MAC computation must be unique for a request to be authenticated.
In particular, this also ensures \textit{policy freshness}, since invocations of \texttt{INSTALL\_POLICY} and \texttt{UPDATE\_POLICY} cannot be replayed to downgrade active policies.

\subsubsection{Policy Confidentiality}


During policy management, important policy-related request data are passed through the host OS to the meta-interface (e.g. the policy code or parameters). However, an adversarial OS may use this information to infer the policy semantics, which will compromise security guarantees for some applications.
This is especially relevant for specific policies, such as those designed for plausibly deniable storage systems~\cite{chen-2021-plausibleDeniability}. 

To prevent the adversary from learning policy semantics when calling into the meta-interface, we introduce a new session mode called the confidentiality mode (in contrast to normal mode). The trusted application enters this mode as part of calling \texttt{INIT\_SESSION}. During this mode, confidentiality is ensured over policy request data for each meta-interface interaction by encrypting request data. To achieve this, the policy confidentiality mode replaces the MAC-based authentication mechanism with an authenticated encryption with associated data (AEAD) scheme, which ensures confidentiality, integrity, and authenticity of request data~\cite{rfc5116}. As the parameters to the authenticated encryption algorithm, the trusted application uses $K_s$ as the key,  $\mathsf{ctr}_1$ as the nonce, the request data as the plaintext, and the NVMe opcode as the associated data. Similar to the normal session mode, the FTL enforces that $\mathsf{ctr}_1 > \mathsf{ctr}_0$ before performing authenticated decryption of the request.

\subsubsection{Policy Protection at Rest}

Calling the meta-interface commands \texttt{INSTALL\_POLICY} and \texttt{UPDATE}\texttt{\_POLICY} stores the plaintext policy code on the SSD, so that they can be easily loaded by the policy engine. To ensure both policy confidentiality and integrity at rest, we store policy code in a reserved on-device region that is not reachable through the host-visible logical address space. This is challenging since policies will present different logical addressing schemes to the host OS, and a policy may map the region containing policy code into the host-visible address space.

To address this, the key idea is to store policy code in a region that is not visible to the policies themselves (i.e., cannot be mapped to eSWDs). To achieve this, policy code is stored within a portion of the SSD's over-provisioned space. Since this space is not mapped to any pseudo-physical block address, it will not be exposed to the host by policies, which only interact with the flash storage through the FTL API. The remaining problem is to prevent policies from eventually mapping this overprovisioned space onto the policy-visible eSWDs. 
To do this, the BBM layer simply refuses to map blocks containing policy code into the policy-managed pseudo-physical address space, preventing this area from being exposed. To manage this storage space, the meta-interface policy uses privileged functions not available to normal policies, interacting directly with the flash storage through the raw flash operations.

\subsubsection{Policy Attestation}

At any point, the system administrator expects a specific set of policies to be installed and active based on the sequence of meta-interface interactions. However an adversarial host OS may manipulate the communications between any trusted application and the meta-interface. Since any synthetic or modified commands will be rejected, the adversary is limited to causing valid requests to become invalid and be rejected, or preventing them from reaching the SSD. While this cannot be prevented under the Dolev-Yao model~\cite{dolev-1983-security}, we enable the system administrator to detect mismatches between the expected and actual on-device policy state.

To support this, we introduce the meta-interface command \texttt{POLICY\_ATTESTATION}, which is a request for the FTL to generate a report on the status of a given policy. The request data for \texttt{POLICY\_ATTESTATION} simply consists of a policy ID, which is maintained by the meta-interface along with other bookkeeping metadata such as the location and size of the policy as part of policy installation. We introduce two report types, depending on the policy properties being audited. 

The first report type is called the security report, and it consists of generating a report over two bits of information, which is sufficient for security. Specifically, the report is generated over (1) a bit indicating whether the policy code associated with the provided ID is installed, and (2) a bit indicating whether the policy is active or not. This is sufficient under our threat model because (i) only authenticated meta-interface commands can install or modify policies, (ii) policy storage is integrity-protected and inaccessible to the host-visible address space. From these, we conclude that policy installation implies policy integrity. Security reports are then generated by computing a MAC or AEAD over the two bits, depending on whether the current session is in normal or confidential mode, respectively.

The second report type is called the consistency report. While the security report is sufficient under our threat model, system-level faults such as firmware bugs or flash corruption may cause policy code to be partially written, lost, or corrupted. Therefore, the consistency report replaces the first bit in the security report with a hash of the relevant policy code. To verify such a report, the trusted application should store a hash of the expected policy code.

\subsection{Policy Engine}
\label{sec:policy-engine}

The policy engine is responsible for ensuring that installed and active policies are enforced by the FTL. When the FTL boots, the $\ours$ firmware is loaded into DRAM and begins execution. However, any ``active'' policies are stored in flash-memory but not yet loaded into system RAM. To do this, the policy engine is invoked on startup, and it references the policy metadata, identifying policies with the active bit set and loading their code into RAM. As established above, (Section~\ref{sec:characterization:model}) policies include special functions that are called when certain ``events'' occur in the FTL. Corresponding to M3, policies can register functions to four event types: The background event, NVMe commands, pSWD state transitions, and primitive flash operation errors. The background event is triggered by the policy engine, NVMe command events by the host OS, and pSWD state transitions and flash operation errors by the flash subsystem. Policies are essentially sets of (condition, action) pairs, implemented in a general purpose programming language, and pointers to these need to be registered with the correct event dispatch table. As a convention, each policy will include a function called $\texttt{INIT\_POLICY}$, which the policy engine calls immediately after loading the policy code into RAM to perform this registration.

During runtime, each event calls a \texttt{DISPATCH} function, which refers to the table associated with that event. Then, for each active entry in the table, \texttt{DISPATCH} will first execute the condition function, then if the condition is satisfied, the action function will be executed. The condition and action functions can call into the FTL API, which is part of the flash memory subsystem, to interact with the storage system.

\subsection{Flash Subsystem}
\label{sec:flash-subsystem}

The flash subsystem is the part of $\ours$ through which policies interact with the flash storage. To distinguish between the device-enforced constraints and policy enforced constraints built on top of them, we split the flash subsystem into three layers, corresponding to M1, M2a, and M2b, respectively: (1) the raw flash operations, (2) the bad block management, and (3) the FTL API.

\subsubsection{Raw Flash Operations.} The lowest layer interacts directly with the flash storage hardware and ensures that all operations satisfy constraints C1-C4. In this layer, the operations are performed directly over physical blocks and pages. Any errors are reported as events to the policy engine. Then, policies handling such events may invoke BBM functions exposed in the FTL API.

\subsubsection{Bad Block Management and pSWDs}

On top of the raw flash operations is the BBM subsystem, which realizes M2a. BBM introduces a pseudo-physical block address space and maintains the mapping from pseudo-physical block addresses to physical blocks. Since each pseudo-physical block is a block-sized striping domain with page-level striping, it is associated with one primitive SWD (pSWD). Thus, this layer maintains the pSWD state machines and context. In addition, BBM provides the functions needed to manage the pseudo-physical to physical mapping, and includes wrappers around the raw flash operations (i.e. pSWD read, write, and erase operations that handle the mapping transparently). Over-provisioned blocks are managed at this layer as physical blocks with no pseudo-physical mapping, which makes them inaccessible through the address space exposed to policies.

BBM also provides a configuration function, \texttt{SET\_OP\_SIZE}, which determines how much physical space is reserved for overprovisioning. This function is exposed to policies through the FTL API and is intended to be invoked by a single designated management policy during \texttt{INIT\_POLICY}. Over-provisioned blocks are distributed evenly across planes to preserve flash-level parallelism after remapping. In addition, the reserved space must also be sufficient to store the policies installed and their metadata.

Finally, BBM provides the basic mechanisms needed for bad-block handling and remapping. When a flash operation error event indicates that a block should be retired, policy can choose to call BBM functions that retire the affected physical block, update the pseudo-physical mapping, and migrate any associated data.

\subsubsection{FTL API and eSWDs}

The top layer of the flash subsystem realizes M2b. It consists of all the functions available to be called directly by policies, as well as the supporting mechanisms required to implement them on top of BBM. Importantly, this layer introduces the eSWD abstraction, which groups pSWDs into policy-visible eSWDs. The FTL API exposes generic read, sequential write, and erase operations over eSWDs, translating each such request into the corresponding BBM operations over the underlying pSWDs.
The device-side eSWD context includes the eSWD size, striping level, per-eSWD write pointer, and the mapping from eSWD onto pSWDs. In particular, the write pointer determines the placement of new writes within the eSWD and enforces the sequential-write requirement.

The FTL API enables policy to determine the eSWD size and striping level via two important functions: 1) \texttt{SET\_ESWD\_SIZE} sets the eSWD size, thereby determining how many eSWDs exist. 2) \texttt{SET\_STRIPING\_LEVEL} sets the striping level, which determines the layout of each eSWD in the pseudo-physical address space, and how its write pointer advances over the associated pSWDs. These functions should be called during \texttt{INIT\_POLICY} by exactly one policy.

The FTL API provides functions for querying and managing the eSWD to pSWD mapping. Also, it supports functions such as wear-leveling, higher-level mappings that enable random writes over a logical address space (such as a standard logical block interface), and garbage collection by providing primitives for migrating eSWDs and abstractions for creating and managing additional mapping tables.

\section{Analysis and Discussion}
\subsection{Security Analysis}
\label{sec:security}

\noindent\textbf{Security Requirements of Extensible SSDs.} Supporting extensibility under a strengthened threat model introduces additional security requirements that do not arise in fixed-function SSDs. In an extensible SSD, a new interface must be introduced through which external entities can influence FTL execution. In $\ours$, this interface is realized as the meta-interface, described in Section~\ref{sec:design:meta}. However, under our threat model (Section~\ref{sec:model}), the meta-interface is always mediated by an untrusted host operating system. The essence of the security challenge is therefore to precisely constrain the conditions under which accessing this interface is allowed to affect FTL execution. Under our threat model, the meta-interface should be accessible only to the trusted system administrator or to applications explicitly delegated by the administrator. Our six security aims are a direct consequence of this observation: (1) Entities attempting to access meta-interface commands must be authenticated, and (2) all legitimate meta-interface interactions must be fresh and non-replayable. This effectively prevents untrusted entities from directly invoking any call into the meta-interface. (3) Meta-interface invocations must be integrity protected such that any modification by the host OS while forwarding requests to the SSD is detectable. (4) Meta-interface invocations must be confidential to prevent the host OS from learning policy semantics. (5) To preserve the confidentiality and integrity of installed policies over time, all interactions with the policy state must be mediated exclusively through the meta-interface. (6) While we cannot prevent DoS attacks under a Dolev-Yao-Style adversary, we must ensure that the meta-interface remains effectively available to the system administrator by making any inconsistencies between the expected and actual system states detectable.

\noindent\textbf{$\ours$ Security Mechanisms.} Our design achieves requirement (1) because the meta-interface verifies either a MAC or AEAD generated using $K_s$ upon every meta-interface invocation. Clearly, an adversary that does not hold $K_s$ cannot generate a valid MAC or AEAD except with negligible probability. Our design achieves requirement (2) because the MAC/AEAD authenticates the session counter $\mathsf{ctr}_1$ and enforces its uniqueness within a session. As a result, any valid MAC/AEAD is either generated under a unique session key or computed over unique data. Requirement (3) follows as a direct consequence of verifying a MAC/AEAD over the request data. $\ours$ achieves requirement (4) by introducing normal and confidential session modes. In normal mode, requests are authenticated using a MAC, whereas in confidential mode, requests are protected using AEAD. We achieve requirement (5) by storing policy to the over-provisioned area, and excluding it from any pseudo-physical block mapping. Since the eSWDs map onto the pseudo-physical address space, no storage interface specified by policy can modify or read the policy storage. In principle, this invariant could be violated if policy code invokes functions outside the FTL API. However, since policy code is provided by a trusted system administrator under our threat model, such violations constitute safety failures rather than security violations. We discuss potential mitigations in \S~\ref{sec:discussion}. Requirement (6) is met by our \texttt{POLICY\_ATTESTATION} meta-interface command, which allows the system administrator to query policy state at any time. Under our threat model, the adversary neither controls an authorized policy manager nor possesses $K_s$, and therefore cannot authenticate policy changes between attestations. Because requirements (1)--(3) and (5) are satisfied, the presence of an installed policy implies that its integrity has been maintained. Therefore, the only state that must be attested is (i) whether a policy is installed, and (ii) whether it is active. Directly attesting policy integrity falls under safety rather than security, since under (1)--(3) and (5), only faulty trusted policy code can corrupt policy storage. Nevertheless, we also support an attestation mode in which policy integrity is explicitly reported.

\subsection{Discussion}
\label{sec:discussion}

\noindent\textbf{Policy Code Safety}. 
As discussed in Section~\ref{sec:security}, we assume that the system administrator is trusted, allowing us to treat policy code installed on the FTL as benign. However, policy code may still violate the development guidelines by invoking functions outside the FTL API, resulting in potential safety failures. Such violations may corrupt policy storage or reveal policy semantics. In addition, policy code may also unintentionally contain vulnerabilities, such as buffer overflows~\cite{aleph1_smashing} or use-after-free errors~\cite{dereference-cwe416}. A promising future direction is to explore the mitigation of these errors through static analysis~\cite{bessey-2010-static}, or restricting policy capabilities during runtime by integrating eBPF into the storage device~\cite{hedam-2024-delilah}.

\noindent\textbf{Policy management by mutually distrusting entities.} Our threat model assumes that all entities authorized to manage policies are mutually trusting. However, there may be scenarios in which a trusted application may need to install a policy while distrusting another authorized policy manager. In this setting, the current \texttt{POLICY\_ATTESTATION} functionality may be insufficient, as only the \textit{current} policy state is audited. A distrusted manager could temporarily replace and later restore the active policy between attestations, violating the security guarantees expected by the trusted application. Supporting this model would require new attestation mechanisms that provide authenticated evidence of policy transitions over time. We leave this to future work. 

\noindent\textbf{Adapting $\ours$ to other SSD interfaces.} Although the $\ours$ meta-interface uses vendor-specific NVMe commands, it can be ported to SATA/SAS using their vendor-specific command mechanisms, without changing the policy engine or flash subsystem significantly. Additionally, policies would have limited capability to create custom vendor-specific commands, as SATA can support 27~\cite{osdev_ata_command_matrix} and SAS can support 64~\cite{t10_scsi_operation_codes} in total.

\section{Implementation and Evaluation}

  \begin{table}[t]
      \centering
      \begin{tabular}{lcc}
        \toprule
        Parameter & Baseline & $\ours$ \\
        \midrule
        Channels & 8 & 8 \\
        LUNs per channel & 8 & 8 \\
        Planes per LUN & 1 & 1 \\
        Blocks per plane & 256 & 276 \\
        Pages per block & 256 & 256 \\
        Page size & 4\,KiB & 4\,KiB \\
        Physical capacity & 16\,GiB & 17.25\,GiB \\
        Policy-visible capacity & 16\,GiB & 16\,GiB \\
        Overprovisioning & 0\% & 7\% \\
        \midrule
        Page read latency & 40\,$\mu$s & 40\,$\mu$s \\
        Page program latency & 200\,$\mu$s & 200\,$\mu$s \\
        Block erase latency & 2\,ms & 2\,ms \\
        \bottomrule
      \end{tabular}
      \caption{Device configuration used in our experiments. Notably, $\ours$ includes internal overprovisioning while keeping the policy-visible
  capacity unchanged.}
 
      \label{tab:device-config}
  \end{table}

Our evaluation is centered around three questions: \textbf{(Q1)} How efficient are the secure policy management functions? \textbf{(Q2)} What is the performance overhead of $\ours$ compared to native FTL implementations? \textbf{(Q3)} Can $\ours$ support different security-critical storage semantics with minimal implementation effort and overhead comparable to native implementations? 

We address \textbf{(Q1)} in Section~\ref{sec:eval:meta-interface}, by measuring the execution time of the vendor-specific NVMe commands introduced by the meta-interface (Table~\ref{tab:meta_nvme_cmds}). We address \textbf{(Q2)} in Section~\ref{sec:eval:baseline-overhead} by implementing a block-interface policy on top of $\ours$ and comparing it against FEMU's native block-interface implementation. To establish a fair comparison, we first demonstrate that the two implementations provide matching semantics by using the traces listed in Table~\ref{tab:workloads} to compare their internal behavior. We then compare throughput and latency to measure the baseline overhead introduced by $\ours$. In Section~\ref{sec:eval:case-studies}, we address \textbf{(Q3)} by implementing FlashGuard~\cite{huang-2017-flashguard}, an FTL-based ransomware recovery strategy, as a modification of our block-interface policy. We also quantify implementation effort by reporting the lines of code required for this modification. We compare the additional overhead introduced by this modification with the overhead reported in the original FlashGuard paper.

\subsection{Implementation}

We have implemented $\ours$ on FEMU~\cite{li-2018-femu}, which is a QEMU-based flash emulator used in many research projects~\cite{han-2021-zns+,zhu-2025-last}.  Our implementation and evaluation artifacts are publicly available~\cite{dafoe2026sxssdartifact} at commit \texttt{3451afc74}. For the baseline comparison, we implemented a block-interface policy that matches the semantics of the original FEMU implementation (we call this $\ours$-Block) in 543 lines of C code. We also implemented a FlashGuard~\cite{huang-2017-flashguard} policy (we call this $\ours$-FlashGuard) as a case study, which is a flash-based ransomware recovery scheme. This policy is built on $\ours$-Block and adds 533 additional lines of code. 
Table~\ref{tab:device-config} shows the configuration of the simulated device. The only configuration difference is that $\ours$ has 7\% extra physical capacity to serve as overprovisioned blocks. A small portion of this space is used to store the policy code (i.e., just 6 pages for $\ours$-Block). Otherwise, the policy-visible and host-visible capacity and geometry are identical to the baseline (i.e., the overprovisioning provides no practical advantage other than providing space for policy code storage). As our evaluation machine, we use a desktop computer with an Intel Core Ultra 7 265K 3.9 GHz and 64GB DDR5 4800 MT/s RAM. To simulate a low-power flash memory controller, we scaled the clock rate to 0.5 GHz from 3.9 GHz by multiplying the \textit{computation} time by 7.8, similar to the methodology used in~\cite{zhu-2025-last}. Finally, the cryptographic primitives were instantiated as follows: the authenticated key exchange for session establishment uses X25519-based Diffie-Hellman and Ed25519 signatures. The key derivation function is HKDF-SHA256, the MAC uses HMAC-SHA256, the hash function is SHA-256, and the authenticated encryption with associated data (AEAD) function uses AES-GCM.

\subsection{Experimental Setup}

  \begin{table}[tb]
\centering
\small
\begin{tabular}{|c|l|c|c|}
  \hline
  \textbf{Family} & \textbf{Workload} & \textbf{ID} & \textbf{Write Ratio} \\
  \hline

  \multirow{3}{*}{Systor~\cite{lee-2017-systor}}
      & Systor LUN 5  & S5  & 49.13\% \\
  \cline{2-4}
      & Systor LUN 6  & S6  & 39.16\% \\
  \cline{2-4}
      & Systor LUN 11 & S11 & 40.19\% \\
  \hline

  \multirow{3}{*}{MSR~\cite{dushyanth-2008-msr}}
      & Proxy server   & M-prxy & 98.36\% \\
  \cline{2-4}
      & Print server   & M-prn  & 95.34\% \\
  \cline{2-4}
      & Media server   & M-mds  & 92.18\% \\
  \hline

  \multirow{3}{*}{FIU~\cite{koller-2010-fiu}}
      & Web server     & F-web2   & 99.98\% \\
  \cline{2-4}
      & Mail server    & F-mail3  & 99.11\% \\
  \cline{2-4}
      & Homes server   & F-homes2 & 99.29\% \\
  \hline

  \multirow{2}{*}{Alibaba~\cite{alibaba-block-traces}}
      & Alibaba device 1 & A1 & 100.00\% \\
  \cline{2-4}
      & Alibaba device 8 & A8 & 100.00\% \\
  \hline
\end{tabular}
\caption{Summary of evaluated workloads.}
\label{tab:workloads}
\end{table}

In our experimental setup, the host operating system is Arch Linux with Linux kernel 6.19.11. To build and launch FEMU, we use an Ubuntu 24.04 container managed by Distrobox with the Docker backend. FEMU runs an Ubuntu 20.04 guest virtual machine with the Linux kernel 5.15.0, configured with 4 GB of RAM and 4 vCPU cores. To avoid performance unpredictability, we set the host CPU to ``performance'' mode, and pin the vCPUs to physical CPUs, using the FEMU-provided scripts~\cite{femu-best-practice}. 

\subsubsection{Workloads and Evaluation Method} In our experiments, we use the FIO benchmarking tool~\cite{fio} to run real-world workload traces and throughput benchmarks. We use workload traces from the SYSTOR'17 dataset~\cite{lee-2017-systor}, Microsoft~\cite{dushyanth-2008-msr}, FIU~\cite{koller-2010-fiu}, and Alibaba~\cite{alibaba-block-traces}. We selected the subset of the trace datasets shown in Table~\ref{tab:workloads} with the highest write ratios. Additionally, we scale the workload request addresses to fit in the host-exposed logical address space, and cap the time between requests at 10ms to save evaluation time. When running the workloads, we split into a warmup phase and an experiment phase: in the warmup phase, we first run FIO random write across 70\% of the logical address space, then run the first 1,000,000 operations of the workload, and for the experiment phase, we capture the timing information during 100,000 additional operations. Before running each workload, we relaunch the VM, then install and activate the relevant policy through the meta-interface (when evaluating $\ours$).

\subsection{Meta-interface Performance}
\label{sec:eval:meta-interface}

  \begin{table}[t]
    \centering
    \setlength{\tabcolsep}{4pt}
    \renewcommand{\arraystretch}{1.15}
    \begin{tabular}{|L{0.50\columnwidth}|C{0.20\columnwidth}|C{0.20\columnwidth}|}
    \hline
    \textbf{Command} & \textbf{Normal (ms)} & \textbf{Confidential (ms)} \\
    \hline
    \begin{filecontents*}{results/meta_interface_results/paper_table.csv}
command,size,normal,confidential
\texttt{INIT\_SESSION},--,1.337,1.132
\texttt{INSTALL\_POLICY},6,0.927,0.920
\texttt{UPDATE\_POLICY},6,1.492,1.468
\texttt{REMOVE\_POLICY},6,0.881,0.896
\texttt{ACTIVATE\_POLICY},6,128.777,132.965
\texttt{DEACTIVATE\_POLICY},6,1.177,1.154
\texttt{POLICY\_ATTESTATION (S)},--,0.223,0.244
\texttt{POLICY\_ATTESTATION (C)},6,0.315,0.344
\end{filecontents*}
\csvreader[
      head to column names,
      late after line=\\\hline
    ]{results/meta_interface_results/paper_table.csv}{}{%
      \command & \normal & \confidential}
    \hline
    \end{tabular}
    \caption{Average device-side execution time of meta-interface commands in normal and confidential modes. The
  \texttt{POLICY\_ATTESTATION} (S) is in security mode, while the \texttt{POLICY\_ATTESTATION} (C) is in
  consistency mode.}
    \label{tab:meta_interface_latency}
  \end{table}

The execution time of the vendor-specific NVMe commands introduced by the meta-interface is shown in Table~\ref{tab:meta_interface_latency}. The evaluation is split into normal mode and confidential mode; overall, the time difference between these modes is negligible. Notably, the consistency-mode attestation has 41\% higher execution time than security mode in both normal and confidential modes, though both operations are still trivially fast. This is expected, since the consistency mode computes a hash over the entire policy code, whereas the security mode only needs to validate two bits of information. By far the most expensive meta-interface operation is \texttt{ACTIVATE\_POLICY}, because it has to 1) read the policy code into memory, 2) parse the policy binary resolve its symbols, and 3) execute the \texttt{INIT\_POLICY} function, which configures the device geometry and loads the policy into the policy engine.

\subsection{Baseline Performance Overhead}
\label{sec:eval:baseline-overhead}

\subsubsection{Baseline and \texorpdfstring{$\ours$}{SxSSD}-Block Policy Semantic Equivalence.}

  \begin{table*}[t]
    \centering
    \small
    \renewcommand{\arraystretch}{0.9}
    \setlength{\tabcolsep}{3pt}
    \setlength{\abovecaptionskip}{3pt}
    \setlength{\belowcaptionskip}{0pt}

    \vspace{-0.75em}
    \begin{adjustbox}{width=\textwidth}
    \begin{filecontents*}{./results/workload-results/csv/semantic_equivalence_percent_diff.csv}
Internal Behavior,S11,S6,S5,M-prxy,M-prn,M-mds,F-web2,F-mail3,F-homes2,A8,A1
Physical page writes,0.000,-0.027,0.000,0.006,0.042,0.006,0.000,0.002,0.000,0.001,0.028
GC invocations,0.000,0.000,0.000,0.000,0.000,0.000,0.000,0.000,0.000,0.000,0.000
GC pages migrated,0.000,-0.342,0.000,4.189,1.174,11.487,0.000,11.444,0.000,0.035,5.219
Block erases,0.000,0.000,0.000,0.000,0.000,0.000,0.000,0.000,0.000,0.000,0.000
\end{filecontents*}
\pgfplotstabletypeset[
        col sep=comma,
        string type,
        every head row/.style={
            before row=\toprule,
            after row=\midrule
        },
        every last row/.style={
            after row=\bottomrule
        },
        every row no 0/.style={before row=\rule{0pt}{2.1ex}},
        columns={Internal Behavior,S11,S6,S5,M-prxy,M-prn,M-mds,F-web2,F-mail3,F-homes2,A8,A1},
        columns/Internal Behavior/.style={column name=\textbf{Internal Behavior}, string type},
        columns/S11/.style={column name=\textbf{S11}},
        columns/S6/.style={column name=\textbf{S6}},
        columns/S5/.style={column name=\textbf{S5}},
        columns/M-prxy/.style={column name=\textbf{M-prxy}},
        columns/M-prn/.style={column name=\textbf{M-prn}},
        columns/M-mds/.style={column name=\textbf{M-mds}},
        columns/F-web2/.style={column name=\textbf{F-web2}},
        columns/F-mail3/.style={column name=\textbf{F-mail3}},
        columns/F-homes2/.style={column name=\textbf{F-homes2}},
        columns/A8/.style={column name=\textbf{A8}},
        columns/A1/.style={column name=\textbf{A1}},
        every column/.style={column type=c},
        columns/Internal Behavior/.append style={column type=l},
    ]{./results/workload-results/csv/semantic_equivalence_percent_diff.csv}
    \end{adjustbox}
    \caption{Semantic equivalence. Each entry reports the percent difference (\%)
  between Baseline and $\ours$-Block.}
    \label{tab:semantic-equivalence}
  \end{table*}

  \begin{table*}[t]
    \centering
    \small
    \renewcommand{\arraystretch}{0.9}
    \setlength{\tabcolsep}{3pt}
    \setlength{\abovecaptionskip}{3pt}
    \setlength{\belowcaptionskip}{0pt}

    \vspace{-0.75em}
    \begin{adjustbox}{width=\textwidth}
    \begin{filecontents*}{./results/workload-results/csv/gc_pages_migrated.csv}
GC Pages Migrated,S11,S6,S5,M-prxy,M-prn,M-mds,F-web2,F-mail3,F-homes2,A8,A1
Baseline,0,585258,4246,1753,100378,919,0,173,0,113314,19278
SxSSD,0,583258,4246,1828,101563,1031,0,194,0,113354,20311
Percent difference,0.000,-0.342,0.000,4.189,1.174,11.487,0.000,11.444,0.000,0.035,5.219
\end{filecontents*}
\pgfplotstabletypeset[
        col sep=comma,
        string type,
        every head row/.style={
            before row=\toprule,
            after row=\midrule
        },
        every last row/.style={
            after row=\bottomrule
        },
        every row no 0/.style={before row=\rule{0pt}{2.1ex}},
        columns={GC Pages Migrated,S11,S6,S5,M-prxy,M-prn,M-mds,F-web2,F-mail3,F-homes2,A8,A1},
        columns/GC Pages Migrated/.style={column name=\textbf{GC Pages Migrated}, string type},
        columns/S11/.style={column name=\textbf{S11}},
        columns/S6/.style={column name=\textbf{S6}},
        columns/S5/.style={column name=\textbf{S5}},
        columns/M-prxy/.style={column name=\textbf{M-prxy}},
        columns/M-prn/.style={column name=\textbf{M-prn}},
        columns/M-mds/.style={column name=\textbf{M-mds}},
        columns/F-web2/.style={column name=\textbf{F-web2}},
        columns/F-mail3/.style={column name=\textbf{F-mail3}},
        columns/F-homes2/.style={column name=\textbf{F-homes2}},
        columns/A8/.style={column name=\textbf{A8}},
        columns/A1/.style={column name=\textbf{A1}},
        every column/.style={column type=c},
        columns/GC Pages Migrated/.append style={column type=l},
    ]{./results/workload-results/csv/gc_pages_migrated.csv}
    \end{adjustbox}
    \caption{GC pages migrated by Baseline and $\ours$.}
    \vspace{-5pt}
    \label{tab:gc-pages-migrated}
  \end{table*}

To ensure that our performance comparison is fair, we first verify that the $\ours$-Block reproduces the same internal behavior as the baseline implementation. Table~\ref{tab:semantic-equivalence} reports the percent difference between the baseline block-interface FTL and $\ours$-Block for several internal behaviors across all workload traces. First, baseline and $\ours$ perform the same number of GC invocations and block erases for every workload, indicating that 1) garbage collection is triggered under the same conditions and 2) the units of garbage collection (i.e., a superblock-sized eSWD) are identical.

Additionally, the physical page write counts differ by at most 0.042\%, indicating negligible differences in write-amplification. Although some workloads show larger percent differences in GC pages migrated in Table~\ref{tab:semantic-equivalence}, Table~\ref{tab:gc-pages-migrated} shows that these occur only when little garbage collection takes place: the average migration-count difference is 0.52\% for workloads with more than 100,000 migrated pages, but 4.04\% for workloads with fewer than 100,000. These differences can be accounted for by the random writes included in the warmup workload causing variations in the overwrite distribution. Overall, these results indicate that the block-interface policy faithfully reproduces the baseline behavior, enabling a fair performance comparison between them. 

\subsubsection{Baseline Overhead.}

\begin{figure}[t]
      \centering
      \begin{tikzpicture}
          \begin{axis}[
              ybar=2.5pt,
              width=0.82\columnwidth,
              height=0.42\columnwidth,
              bar width=5.0pt,
              ylabel={Mean write latency ($\mu$s)},
              symbolic x coords={S11,S6,S5,M-prxy,M-prn,M-mds,F-web2,F-mail3,F-homes2,A8,A1},
              xtick=data,
              x tick label style={rotate=45, anchor=east, font=\scriptsize},
              ymajorgrids=true,
              grid style={dashed,gray!30},
              ymin=0,
              enlarge x limits=0.06,
              legend style={
                  at={(0.5,1.03)},
                  anchor=south,
                  legend columns=3,
                  draw=none,
                  font=\small
              },
              legend cell align={left},
              area legend,
              tick label style={font=\small},
              label style={font=\small},
          ]
              \addplot[
                  draw=black,
                  fill=nord9,
                  postaction={pattern=north east lines}
              ] table[
                  x=Workload,
                  y={Baseline (us)},
                  col sep=comma
              ] {./results/workload-results/csv/write_clat_mean_us_flashguard.csv};

              \addplot[
                  draw=black,
                  fill=nord14,
                  postaction={pattern=dots}
              ] table[
                  x=Workload,
                  y={SxSSD-Block (us)},
                  col sep=comma
              ] {./results/workload-results/csv/write_clat_mean_us_flashguard.csv};

              \addplot[
                  draw=black,
                  fill=nord11,
                  postaction={pattern=crosshatch}
              ] table[
                  x=Workload,
                  y={SxSSD-FlashGuard (us)},
                  col sep=comma
              ] {./results/workload-results/csv/write_clat_mean_us_flashguard.csv};

              \legend{Baseline,$\ours$-Block,$\ours$-FlashGuard}
          \end{axis}
      \end{tikzpicture}
      \vspace{-10pt}
      \caption{Mean write completion latency across evaluated workloads for baseline, $\ours$-Block, and $\ours$-FlashGuard.}
      \label{fig:write-clat}
  \end{figure}
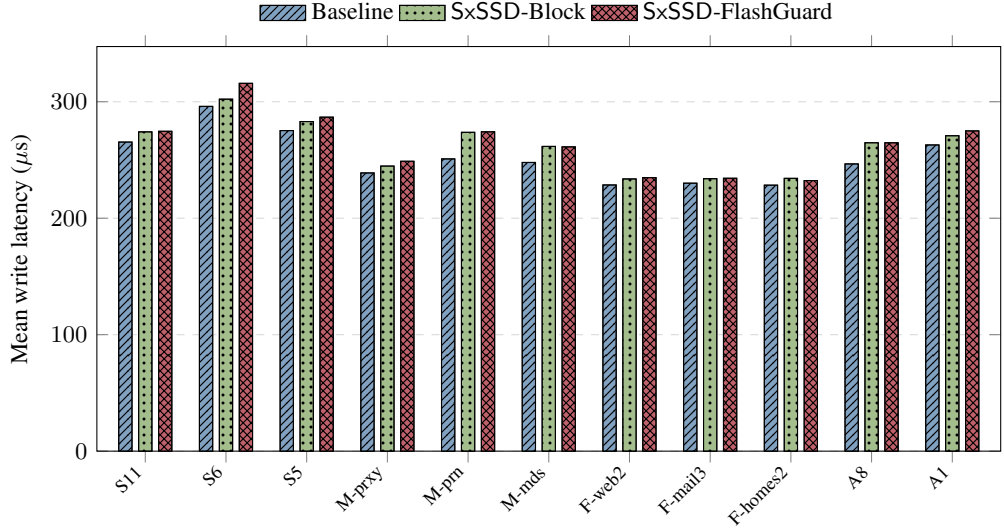

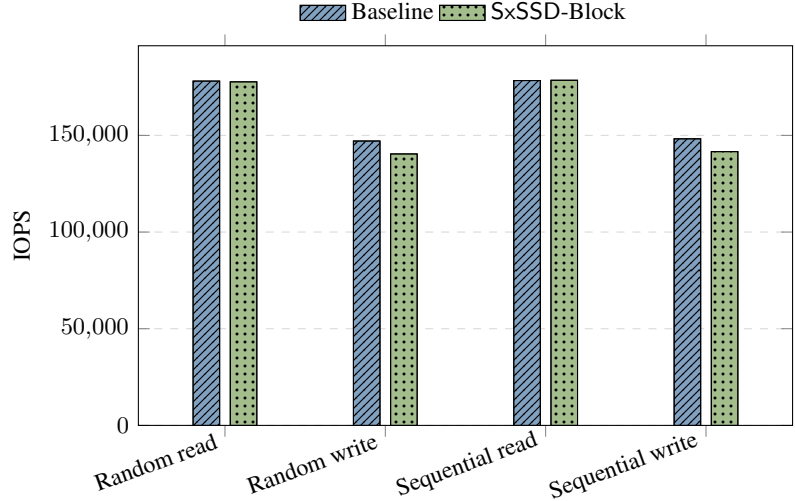
\begin{figure}[t]
      \centering
      \begin{tikzpicture}
          \begin{axis}[
              ybar=4pt,
              width=0.62\columnwidth,
              height=0.40\columnwidth,
              bar width=10pt,
              ylabel={IOPS},
              symbolic x coords={Random read,Random write,Sequential read,Sequential write},
              xtick=data,
              x tick label style={font=\small, rotate=20, anchor=east},
              ymajorgrids=true,
              grid style={dashed,gray!30},
              ymin=0,
              enlarge x limits=0.18,
              legend style={
                  at={(0.5,1.03)},
                  anchor=south,
                  legend columns=2,
                  draw=none,
                  font=\small
              },
              legend cell align={left},
              area legend,
              tick label style={font=\small},
              label style={font=\small},
              scaled y ticks=false,
              yticklabel={\pgfmathprintnumber[fixed,precision=0]{\tick}},
          ]
              \addplot[
                  draw=black,
                  fill=nord9,
                  postaction={pattern=north east lines}
              ] table[
                  x=Workload,
                  y={Baseline (IOPS)},
                  col sep=comma
              ] {./results/workload-results/csv/peak_rand_iops.csv};

              \addplot[
                  draw=black,
                  fill=nord14,
                  postaction={pattern=dots}
              ] table[
                  x=Workload,
                  y={SxSSD (IOPS)},
                  col sep=comma
              ] {./results/workload-results/csv/peak_rand_iops.csv};

              \legend{Baseline,$\ours$-Block}
          \end{axis}
      \end{tikzpicture}
      \caption{Peak random and sequential read/write IOPS under high-intensity 4 KiB fio workloads.}
      \label{fig:peak-rand-iops}
  \end{figure}

Next, we compare the performance of the baseline implementation, $\ours$-Block. Since the policy semantics are the same, this captures a reliable measure of the overhead introduced by $\ours$ overall. Figure~\ref{fig:write-clat} shows the average latency to complete a single write operation for each workload trace. Across all workloads, the average latency introduced by $\ours$ is only 3.74\%. Corresponding to this, Figure~\ref{fig:peak-rand-iops} shows that throughput is also slightly impacted. While random-read IOPS decreases by only  0.069\%, random-write IOPS decrease by 4.617\% relative to the baseline. These results indicate that $\ours$ imposes negligible overhead on reads and a small overhead on writes.

\subsection{Case Study}
\label{sec:eval:case-studies}

\subsubsection{Recovery From Ransomware Attacks.}
FlashGuard was the first work to introduce data recovery from malware attacks leveraging the out-of-place update behavior of block-interface FTLs~\cite{huang-2017-flashguard}. It tracks recently read logical pages, and if such a page is later overwritten, FlashGuard treats the old version as suspicious and preserves it during garbage collection for recovery after an attack. In order to later restore retained pages after an attack, FlashGuard stores metadata such as the associated logical page address and previous (pseudo-)physical page address in the per-page OOB area.

We implemented $\ours$-FlashGuard as a modification of $\ours$-Block. Notably, we achieved this with only 533 additional lines of policy code, and without changing the FTL itself. The logic for retaining the previously read pages is implemented in the policy code attached to existing $\ours$ events. In particular, the read behavior is implemented by modifying the function attached to the read NVMe command event. This change is to mark the corresponding bit in a read bitmap before passing control to the normal block-interface read function.
The write behavior is implemented similarly by modifying the function attached to the write NVMe command event. When a logical page is written, it first checks whether the page has been read. If so, the policy marks the associated pseudo-physical page as retained. Otherwise, the page is invalidated normally. In addition, every write adds the necessary OOB metadata before allocating and writing to a new pseudo-physical page. The garbage collection, which is registered to the background event, is modified by migrating retained pages in addition to valid pages, and copying the OOB along with migrated pages. Additionally, the FlashGuard recovery interface is exposed through two policy-defined vendor specific NVMe commands.
Overall, this demonstrates that FlashGuard is naturally expressible as a policy on $\ours$.

Next, we briefly evaluate the additional overhead introduced by the FlashGuard policy. We ran all workloads on $\ours$-FlashGuard, and the results are displayed in Figure~\ref{fig:write-clat}. Relative to $\ours$-Block, the mean write completion latency increased by 4.50\% at most, for the S6 workload. However, the average increase in latency across all workloads is only 0.86\%. This difference arises because most of the workloads lack 1) sufficient read requests and 2) sufficient garbage collection. In the FlashGuard paper, the average latency increases by at most 6.1\%~\cite{huang-2017-flashguard}, which is consistent with our results.

\section{Related Work}

\noindent\textbf{Software-Defined and Extensible SSDs.}
Ouyang et al.~\cite{ouyang-2014-sdf} and Zhang et al.~\cite{zhang-2016-SDF} were among the first to introduce software-defined SSDs. Their strategy was for the SSD to expose an interface close to the flash hardware characteristics and move FTL functions into the host OS. This work is a conceptual precursor to LightNVM~\cite{bjorling-2017-lightNVM}, which introduced a Linux subsystem for open-channel SSD management. As a middle ground between fully host-managed OCSSDs and traditional block-interface SSDs, Zoned Namespaces SSDs were subsequently introduced~\cite{bjorling-2021-zns,nvme_zns_spec_1_3}. Seshadri et al. proposed Willow~\cite{seshadri-2014-willow}, an extensible SSD that enables device-enforced application-defined storage semantics through SSD Apps (e.g., Base-IO or Atomic-Writes). However, their design does not consider NAND flash and rather uses simulated phase change memory, which is not restricted by C1-C4. As a result, Willow adds new application-level operations to the SSD, rather than making the NAND flash FTL itself programmable. Additionally, Willow does not consider a compromised OS and therefore does not support security-critical storage semantics.

\noindent\textbf{FTL Based Security Critical Storage Semantics.}
Previous work has demonstrated that several security critical applications can be supported by extending the FTL with special storage semantics. FlashGuard~\cite{huang-2017-flashguard} and SSD-Insider~\cite{baek-2018-ssdInsider} enable recovery from ransomware attacks by taking advantage of the out-of-place updates inherent in block-interface FTLs. PEARL enables plausible deniability by storing hidden bits in the same physical locations as public bits~\cite{chen-2021-plausibleDeniability}. HiDCS modifies the FTL to enable self-auditing and self-repair for storage peers in the decentralized cloud, ensuring data integrity without relying on expensive consensus mechanisms~\cite{dafoe-2025-hardware}. 

\noindent\textbf{Generic FTL Frameworks.}
HIL is a conceptual framework for FTL development that uses Logs as the fundamental building block~\cite{choi-2018-hil}. It provides a flexible multi-layer mapping framework that also supports crash recovery. OX establishes a framework for designing custom FTLs on top of Open-Channel SSDs by deconstructing the FTL into modular components and defining their interactions~\cite{picoli2019ox}.

\section{Conclusion}

In this work, we have introduced $\ours$, the first secure yet extensible software-defined SSD. $\ours$ enables restricted access for trusted applications to program the FTL behavior by developing and installing custom policies. In addition, $\ours$ is NAND flash aware and addresses the limitations of previous work by making the flash-to-OS interface changeable. Since the FTL is running on isolated flash memory hardware, it constitutes a critical security boundary. Thus, $\ours$ supports the development and dynamic installation of security-critical storage semantics even when the operating system is compromised. To demonstrate this, we have implemented a case study in which we enable secure data recovery from ransomware attacks. In addition, we evaluate $\ours$ across several real-world workloads.

\section*{Acknowledgment}
This material is based upon work supported by the National Science Foundation Graduate Research Fellowship Program under Grant No. 2437847, and by the National Science Foundation under Grant Nos. 2225424, 2043022 and 2623031. Any opinions, findings, and conclusions or recommendations expressed in this material are those of the author(s) and do not necessarily reflect the views of the National Science Foundation.

\bibliographystyle{unsrt}
\bibliography{refs}


\end{document}